\documentclass[
nojss]{jss}

\usepackage{orcidlink,thumbpdf,lmodern}

\usepackage[utf8]{inputenc}

\author{
Piotr Chlebicki~\orcidlink{0009-0006-4867-7434}\\Stockholm
University \And Maciej
Beręsewicz~\orcidlink{0000-0002-8281-4301}\\Poznań University of
Economics and Business\\
Statistical Office in Poznań
}
\title{\pkg{singleRcapture}: An \proglang{R} Package for Single-Source
Capture-Recapture Models}

\Plainauthor{Piotr Chlebicki, Maciej Beręsewicz}
\Plaintitle{singleRcapture: An R Package for Single-Source
Capture-Recapture Models}
\Shorttitle{\pkg{singleRcapture}: Single-Source Capture-Recapture
Models}

\Abstract{
Population size estimation is a major challenge in official statistics,
social sciences, and natural sciences. The problem can be tackled by
applying capture-recapture methods, which vary depending on the number
of sources used, particularly on whether a single or multiple sources
are involved. This paper focuses on the first group of methods and
introduces a novel \proglang{R} package: \pkg{singleRcapture}. The
package implements state-of-the-art single-source capture-recapture
(SSCR) models (e.g.~zero-truncated one-inflated regression) together
with new developments proposed by the authors, and provides a
user-friendly application programming interface (API). This
self-contained package can be used to produce point estimates and their
variance and implements several bootstrap variance estimators or
diagnostics to assess quality and conduct sensitivity analysis. It is
intended for users interested in estimating the size of populations,
particularly those that are difficult to reach or measure, for which
information is available only from one source and dual/multiple system
estimation is not applicable. Our package serves to bridge a significant
gap, as the SSCR methods are either not available at all or are only
partially implemented in existing R packages and other open-source
software.
}

\Plainkeywords{population size estimation, hidden populations, truncated
distributuons, count regression models, R}

\Address{
    Piotr Chlebicki\\
    Stockholm University\\
    Matematiska institutionen\\
Albano hus 1\\
106 91 Stockholm, Sweden\\
  E-mail: \email{piotr.chlebicki@math.su.se}\\
  URL: \url{https://github.com/Kertoo},
\url{https://www.su.se/profiles/pich3772}\\~\\
      Maciej Beręsewicz\\
    Poznań University of Economics and Business\\
Statistical Office in Poznań\\
    \hfill\break
Poznań University of Economics and Business\\
Department of Statistics\\
Institute of Informatics and Quantitative Economics\\
Al. Niepodległosci 10\\
61-875 Poznań, Poland\\
\strut \\
Statistical Office in Poznań\\
ul. Wojska Polskiego 27/29\\
60-624 Poznań, Poland\\
  E-mail: \email{maciej.beresewicz@ue.poznan.pl}\\
  URL: \url{https://github.com/BERENZ},
\url{https://ue.poznan.pl/en/people/dr-maciej-beresewicz/}\\~\\
  }

\providecommand{\tightlist}{%
  \setlength{\itemsep}{0pt}\setlength{\parskip}{0pt}}

\usepackage{amsmath, amsthm, amssymb} \usepackage{calc, ragged2e} \usepackage[ruled]{algorithm2e}

\DeclareMathAlphabet{\mathmybb}{U}{bbold}{m}{n}

\begin{document}

\section{Introduction}\label{sec-introduction}

Population size estimation is a methodological approach employed across
multiple scientific disciplines, which serves as the basis for research,
policy formulation, and decision-making processes
\citep[cf.][]{bohning2018capture}. In the field of statistics,
particularly official statistics, precise population estimates are
essential in order to develop robust economic models, optimize resource
allocation, and inform evidence-based policy
\citep[cf.][]{baffour2013modern}. Social scientists utilize advanced
population estimation techniques to investigate \emph{hard-to-reach}
populations, such as homeless individuals or illicit drug users in an
effort to overcome the inherent limitations of conventional census
methodologies. These techniques are crucial for obtaining accurate data
on populations that are typically under-represented or difficult to
access using traditional sampling methods
\citep[cf.][]{vincent2022estimating}. In ecology and epidemiology,
researchers focus on estimating the size of individual species or
disease-affected populations within defined geographical regions as part
of conservation efforts, ecosystem management, and public health
interventions.

Population size estimation can be approached using various
methodologies, each with distinct advantages and limitations.
Traditional approaches include full enumeration (e.g.~censuses) and
comprehensive sample surveys, which, while providing detailed data, are
often resource-intensive and may result in delayed estimates,
particularly for human populations. Alternative methods leverage
existing data sources, such as administrative registers or carefully
designed small-scale studies in wildlife research, or census coverage
surveys \citep[cf.][]{wolter1986some, zhang2019note}. Information from
these sources is often extracted by applying statistical methods, known
as \emph{capture-recapture} or \emph{multiple system estimation}, which
rely on data from multiple enumerations of the same population
\citep[cf.][]{dunne2024system}. This approach can be implemented using
either a single source with repeated observations, two sources, or
multiple sources.

In this paper we focus on methods that involve a single data source with
multiple enumerations of the same units \citep[cf.][]{ztpoisson}. In
human population studies, such data can be derived from police records,
health system databases, or border control logs; in the case of
non-human populations, data of this kind can come from veterinary
records or specialized field data. These methods are often applied to
estimate hard-to-reach or hidden populations, where standard sampling
methods may be inappropriate because of prohibitive costs or problems
with identifying population members.

While methods for two or more sources are implemented in various
open-source software packages, for instance \pkg{CARE-4}
\citep{yang2006care4} (in \proglang{GAUSS}), \pkg{Rcapture}
\citep{baillargeon2007rcapture}, \pkg{marked} \citep{laake2013marked} or
\pkg{VGAM} \citep{yee2015vgam} (all in \proglang{R}), single-source
capture-recapture (SSCR) methods are either not available at all or are
only partially implemented in existing \proglang{R} packages or other
software. Therefore, the paper attempts to bridge this gap by
introducing the \pkg{singleRcapture} package, which implement
\emph{state-of-the-art} SSCR methods and offer a user friendly API
resembling existing \proglang{R} functions (e.g., \texttt{glm}). In the
next subsection we describe existing \proglang{R} packages and other
software that could be used for estimating population size using SSCR
methods.

\subsection{Software for capture-recapture with a single
source}\label{sec-software}

The majority of SSCR methods assume zero-truncated distributions or
their extensions (e.g., inclusion of one-inflation). The
\pkg{Distributions.jl} \citep{Distributionsjl} (in \proglang{Julia}),
\pkg{StatsModels} \citep{seabold2010statsmodels} (in \proglang{Python}),
\pkg{countreg} \citep{countreg}, \pkg{VGAM} \citep{VGAM-main} or
\pkg{distributions3} \citep{distributions3} (in \proglang{R}) implement
some of those truncated distributions
(e.g.~\code{distributions3::ZTPoisson} or \code{countreg::zerotrunc})
and the most general distributions, such as Generally Altered, Inflated,
Truncated and Deflated, can be found in the \pkg{VGAM} package
(e.g.~\code{VGAM::gaitdpoisson} for the Poisson distribution, see
\citet{gaitdcount} for a detailed description). However, the estimation
of parameters of a given truncated (and possibly inflated) distribution
is just the first step (as in the case of log-linear models in
capture-recapture with two sources) and, to the best of our knowledge,
there is no open-source software that can be used to estimate population
size using on SSCR methods and includes variance estimators or
diagnostics.

Therefore, the purpose of the \pkg{singleRcapture}, an \proglang{R}
package, is to bridge this gap by providing scientists and other
practitioners with a tool for estimating population size with SSCR
methods. We have implemented state-of-the-art methods, as recently
described by \citet{bohning2018capture} or \citet{bohning2024one} and
provided their extensions (e.g., inclusion of covariates, different
treatment of one-inflation), which will be covered in detail in Section
\ref{sec-theory}. The package implements variance estimation based on
various methods, can be used to create custom models and diagnostic
plots (e.g.~rootograms) with parameters estimated using a modified
iteratively reweighted least squares (IRLS) algorithm we have
implemented for estimation stability. To the best of our knowledge, no
existing open-source package allows the estimation of population size by
selecting an appropriate SSCR model, conducting the estimation, and
providing informative diagnostics of the results.

The remaining part of the paper is structured as follows. Section
\ref{sec-theory} contains a brief description of the theoretical
background and information about fitting methods and available methods
of variance estimation. Section \ref{sec-main} introduces the main
functions of the package. Section \ref{sec-study} presents a case study
and contains an assessment of its results, diagnostics and estimates of
specific sub-populations. Section \ref{sec-methods} describes classes
and \code{S3methods} implemented in the package. The paper ends with
conclusions and an appendix showing how to use the
\code{estimatePopsizeFit} function which is aimed to mimic the
\code{glm.fit} or similar functions. In replication materials we show
how to implement a custom model as this option could be of interest to
users wishing to apply any new bootstrap methods not implemented in the
package.

\section{Theoretical background}\label{sec-theory}

\subsection{How to estimate population size with a single
source?}\label{how-to-estimate-population-size-with-a-single-source}

Let \(Y_{k}\) represent the number of times the \(k\)-th unit was
observed in a single source (e.g.~register). Clearly, we only observe
\(k:Y_{k}>0\) and do not know how many units have been missed
(i.e.~\(Y_{k}=0\)), so the population size, denoted by \(N\), needs to
be estimated. In general, we assume that the conditional distribution of
\(Y_{k}\) given a vector of covariates \(\boldsymbol{x}_{k}\) follows a
version of the zero-truncated count data distribution (and its
extensions). When we know the parameters of the distribution we can
estimate the population size using a Horvitz-Thompson type estimator
given by:

\begin{equation}
\hat{N}=
\sum_{k=1}^{N}\frac{I_{k}}{\mathbb{P}[Y_{k}>0|\boldsymbol{X}_{k}]}=
\sum_{k=1}^{N_{obs}}\frac{1}{\mathbb{P}[Y_{k}>0|\boldsymbol{X}_{k}]},
\label{eq-ht-estimator}
\end{equation}

where \(I_{k}:=\mathcal{I}_{\mathbb{N}}(Y_{k})\), \(N_{obs}\) is the
number of observed units and \(\mathcal{I}\) is the indicator function,
while the maximum likelihood estimate of \(N\) is obtained after
substituting regression parameters \(\boldsymbol{\beta}\) for
\(\mathbb{P}[Y_{k}>0|\boldsymbol{x}_{k}]\) in \eqref{eq-ht-estimator}.

The basic SSCR assumes independence between counts, which is a rather
naive assumption, since the first capture may significantly influence
the behavior of a given unit or limit the possibility of subsequent
captures (e.g.~due to incarceration).

To solve these issues, \citet{godwin2017estimation} and
\citet{ztoi-oizt-poisson} introduced one-inflated distributions, which
explicitly model the probability of singletons by giving additional mass
\(\omega\) to singletons denoted as \(\mathcal{I}_{\{1\}}(y)\)
\citep[cf.][]{bohning2024one}. In other words they considered a new
random varialbe \(Y^{\ast}\) that corresponds to the data collection
process which exhibits one inflation:

\begin{equation*}
  \mathbb{P}\left[Y^{\ast}=y|Y^{\ast}>0\right] =
  \omega\mathcal{I}_{\{1\}}(y)+(1-\omega)\mathbb{P}[Y=y|Y>0].
\end{equation*}

Analytic variance estimation is then performed by computing two parts of
the decomposition according to the law of total variance given by:

\begin{equation}\label{eq-law_of_total_variance_decomposition}
  \text{var}[\hat{N}] = \mathbb{E}\left[\text{var}
  \left[\hat{N}|I_{1},\dots,I_{n}\right]\right] +
  \text{var}\left[\mathbb{E}[\hat{N}|I_{1},\dots,I_{n}]\right],
\end{equation}

where the first part can be estimated using the multivariate \(\delta\)
method given by:

\begin{equation*}
  \mathbb{E}\left[\text{var} \left[\hat{N}|I_{1},\dots,I_{n}\right]\right] =
  \left.\left(\frac{\partial(N|I_1,\dots,I_N)}{\partial\boldsymbol{\beta}}\right)^\top
  \text{cov}\left[\hat{\boldsymbol{\beta}}\right]
  \left(\frac{\partial(N|I_1,\dots,I_N)}{\partial\boldsymbol{\beta}}\right)
  \right|_{\boldsymbol{\beta}=\hat{\boldsymbol{\beta}}},
\end{equation*}

while the second part of the decomposition in
\eqref{eq-law_of_total_variance_decomposition}, assuming independence of
\(I_{k}\)'s and after some omitted simplifications, is optimally
estimated by:

\begin{equation*}
  \text{var}\left[\mathbb{E}(\hat{N}|I_{1},\dots,I_{n})\right] =
  \text{var}\left[\sum_{k=1}^{N}\frac{I_{k}}{\mathbb{P}(Y_{k}>0)}\right]
  \approx\sum_{k=1}^{N_{obs}}\frac{1-\mathbb{P}(Y_{k}>0)}{\mathbb{P}(Y_{k}>0)^{2}},
\end{equation*}

which serves as the basis for interval estimation. Confidence intervals
are usually constructed under the assumption of (asymptotic) normality
of \(\hat{N}\) or asymptotic normality of \(\ln(\hat{N}-N)\) (or log
normality of \(\hat{N}\)). The latter is an attempt to address a common
criticism of student type confidence intervals in SSCR, namely a
possibly skewed distribution of \(\hat{N}\), and results in the
\(1-\alpha\) confidence interval given by:

\begin{equation*}
  \left(N_{obs}+\frac{\hat{N}-N_{obs}}{\xi},N_{obs} +
  \left(\hat{N}-N_{obs}\right)\xi\right),
\end{equation*}

where:

\begin{equation*}
  \xi = \exp\left(z\left(1-\frac{\alpha}{2}\right)
  \sqrt{\ln\left(1+\frac{\widehat{\text{Var}}(\hat{N})}{\left(\hat{N}-N_{obs}\right)^{2}}\right)}\right),
\end{equation*}

and where \(z\) is the quantile function of the standard normal
distribution. The estimator \(\hat{N}\) is best interpreted as being an
estimator of the total number of \underline{observable} units in the
population, since we have no means of estimating the number of units in
the population for which the probability of being included in the data
is \(0\) \citep[cf.][]{ztpoisson}.

\subsection{Available models}\label{available-models}

The full list of models implemented in \pkg{singleRcapture} along with
corresponding expressions for probability density functions and point
estimates can be found in the collective help file for all family
functions:

\begin{CodeChunk}
\begin{CodeInput}
R> ?ztpoisson
\end{CodeInput}
\end{CodeChunk}

For the sake of simplicity, we only list the family functions together
with brief descriptions. For more detailed information, please consult
the relevant literature.

The current list of these family functions includes:

\begin{itemize}
    \item Generalized Chao's \citep{chao1987estimating} and Zelterman's \citep{zelterman1988robust} estimators via logistic regression on variable $Z$ defined as $Z=1$ if $Y=2$ and $Z=0$ if $Y=1$ with $Z\sim b(p)$ where $b(\cdot)$ is the Bernoulli distribution and $p$ can be modeled for each unit $k$ by $\text{logit}(p_k)=\ln(\lambda_k/2)$ with Poisson parameter $\lambda_k=\boldsymbol{x}_{k}\boldsymbol{\beta}$ (for a covariate extension see \cite{chao-generalization} and \cite{zelterman}):
    \begin{align}
        \hat{N}_{\text{Chao}} &= N_{obs}+
        \sum_{k=1}^{\boldsymbol{f}_{1}+\boldsymbol{f}_{2}}
        \left(2\exp\left(\boldsymbol{x}_{k}\hat{\boldsymbol{\beta}}\right)+
        2\exp\left(2\boldsymbol{x}_{k}\hat{\boldsymbol{\beta}}\right)\right)^{-1},
        \tag{\text{Chao's estimator}}\\
        \hat{N}_{\text{Zelt}}&=\sum_{k=1}^{N_{obs}}
        \left(1-\exp\left(-2\exp\left(\boldsymbol{x}_{k}\hat{\boldsymbol{\beta}}\right)\right)\right)^{-1}.
        \tag{\text{Zelterman's estimator}}
    \end{align}
    where $\boldsymbol{f}_{1}$ and $\boldsymbol{f}_{2}$ denotes number of units observed once and twice.
    \item Zero-truncated (\code{zt}$^\ast$) and zero-one-truncated (\code{zot}$^\ast$) Poisson \citep[cf. ][]{zotmodels}, geometric, NB type II (NB2) regression, where the non-truncated distribution is parameterized as:
    \begin{equation*}
        \mathbb{P}[Y=y|\lambda,\alpha] = \frac{\Gamma\left(y+\alpha^{-1}\right)}{\Gamma\left(\alpha^{-1}\right)y!}
        \left(\frac{\alpha^{-1}}{\alpha^{-1}+\lambda}\right)^{\alpha^{-1}}
        \left(\frac{\lambda}{\lambda + \alpha^{-1}}\right)^{y}.
    \end{equation*}
    \item Zero-truncated one-inflated (\code{ztoi}$^\ast$) modifications, where the count data variable $Y^{\ast}$ is defined such that its distribution statisfies:
    \begin{equation*}
    \mathbb{P}\left[Y^{\ast}=y\right]=
    \begin{cases}
    \mathbb{P}[Y=0] & y=0, \\
    \omega\left(1-\mathbb{P}[Y=0]\right)+(1-\omega)\mathbb{P}[Y=1] & y=1, \\
    (1-\omega)\mathbb{P}[Y=y] & y>1,
    \end{cases}
    \end{equation*}
    \begin{equation*}
        \mathbb{P}\left[Y^{\ast}=y|Y^{\ast}>0\right]=\omega\mathcal{I}_{\{1\}}(y)+(1-\omega)\mathbb{P}[Y=y|Y>0].
    \end{equation*}
    \item One-inflated zero-truncated (\code{oizt}$^\ast$) modifications, where the count data variable $Y^{\ast}$ is defined as:
    \begin{equation*}
        \mathbb{P}\left[Y^{\ast}=y\right] = \omega \mathcal{I}_{\{1\}}(y)+(1-\omega)\mathbb{P}[Y=y],
    \end{equation*}
    \begin{equation*}
        \mathbb{P}\left[Y^{\ast}=y|Y^{\ast}>0\right] =
        \omega\frac{\mathcal{I}_{\{1\}}(y)}{1-(1-\omega)\mathbb{P}[Y=0]}+
        (1-\omega)\frac{\mathbb{P}[Y=y]}{1-(1-\omega)\mathbb{P}[Y=0]}.
    \end{equation*}
    Note that \code{ztoi}$^\ast$ and \code{oizt}$^\ast$ distributions are equivalent, in the sense that the maximum value of the likelihood function is equal for both of those distributions given any data, as shown by \cite{bohning2023equivalence} but population size estimators are different.
\end{itemize}

In addition, we propose two new approaches to model singletons in a
similar way as in hurdle models. In particular, we have proposed the
following:

\begin{itemize}
    \item The zero-truncated hurdle model (\code{ztHurdle}$^\ast$) for Poisson, geometric and NB2 is defined as:
    \begin{equation*}
        \mathbb{P}\left[Y^{\ast}=y\right]=\begin{cases}
        \frac{\mathbb{P}[Y=0]}{1-\mathbb{P}[Y=1]} & y=0, \\
        \pi(1-\mathbb{P}[Y=1]) & y=1, \\
        (1-\pi) \frac{\mathbb{P}[Y=y]}{1-\mathbb{P}[Y=1]} & y>1,
        \end{cases}
    \end{equation*}
    \begin{equation*}
        \mathbb{P}\left[Y^{\ast}=y|Y^{\ast}>0\right]=\pi\mathcal{I}_{\{1\}}(y)+
        (1-\pi)\mathcal{I}_{\mathbb{N}\setminus\{1\}}(y)\frac{\mathbb{P}[Y=y]}{1-\mathbb{P}[Y=0]-\mathbb{P}[Y=1]}.
    \end{equation*}
    where $\pi$ denotes the conditional probability of observing singletons.
    \item The hurdle zero-truncated model (\code{Hurdlezt}$^\ast$) for Poisson, geometric and NB2 is defined as:
    \begin{align*}
        \mathbb{P}\left[Y^{\ast}=y\right]&=\begin{cases}
        \pi & y=1, \\
        (1-\pi) \frac{\mathbb{P}[Y=y]}{1-\mathbb{P}[Y=1]} & y\neq1,
        \end{cases}\\
        \mathbb{P}\left[Y^{\ast}=y|Y^{\ast}>0\right]&=\begin{cases}
            \pi\frac{1-\mathbb{P}[Y=1]}{1-\mathbb{P}[Y=0]-\mathbb{P}[Y=1]} & y=1,\\
            (1-\pi)\frac{\mathbb{P}[Y=y]}{1-\mathbb{P}[Y=0]-\mathbb{P}[Y=1]} & y>1,
        \end{cases}
    \end{align*}
    where $\pi$ denotes the unconditional probability of observing singletons.
\end{itemize}

The approaches presented above differ in their assumptions,
computational complexity, or in the way they treat heterogeneity of
captures and singletons. For instance, the dispersion parameter
\(\alpha\) in the NB2 type models is often interpreted as measuring the
\textit{severity} of unobserved heterogeneity in the underlying Poisson
process \citep[cf.][]{ztnegbin}. When using any truncated NB model, the
hope is that given the class of models considered, the consistency is
not lost despite the lack of information.

While not discussed in the literature, the interpretation of
heterogeneous \(\alpha\) across the population (specified in
\code{controlModel}) would be that the unobserved heterogeneity affects
the accuracy of the prediction for the dependent variable \(Y\) more
severely than others. The geometric model (NB with \(\alpha=1\)) is
singled out in the package and often considered in the literature
because of inherent computational issues with NB models, which are
exacerbated by the fact that data used for SSCR are usually of rather
low quality. Data sparsity is a particularly common problem in SSCR and
a big challenge for all numerical methods for fitting the
(zero-truncated) NB model.

The extra mass \(\omega\) in one-inflated models is an important
extension to the researcher's toolbox for SSCR models, since the
inflation at \(y=1\) is likely to occur in many types of applications.
For example, when estimating the number of active people who committed
criminal acts in a given time period, the fact of being captured for the
first time following an arrest is associated with the risk of no longer
being able to be captured a second time. One constraint present in
modelling via inflated models is that attempts to include both the
possibility of one inflation and one deflation lead to both numerical
and inferential problems since the parameter space (of
\((\omega, \lambda)\) or \((\omega, \lambda, \alpha)\)) is then given by
\(\{(\omega, \lambda, \alpha) | \forall x\in \mathbb{N}: p(x|\omega, \lambda, \alpha)\geq0\}\)
for the probability mass function \(p\). The boundary of this set is
then a \(1\) or \(2-\)dimentional manifold, transforming this parameter
space into \(\mathbb{R}^{3}\) would require using ``link'' functions
that depend on more than one parameter.

Hurdle models represent another approach to modelling one-inflation.
They can also model deflation as well as inflation and deflation
simultaneously, so they are more flexible and, in the case of hurdle
zero-truncated models, appear to be more numerically stable.

Although the question of how to interpret regression parameters tends to
be somewhat overlooked in SSCR studies, we should point out that the
interpretation of the \(\omega\) inflation parameter (in
\code{ztoi}\(^\ast\) or \code{oizt}\(^\ast\)) is more convenient than
the interpretation of the \(\pi\) probability parameter (in hurdle
models). Additionally, the interpretation of the \(\lambda\) parameter
in (one) inflated models conforms to the following intuition: given that
unit \(k\) comes from the non-inflated part of the population, it
follows a Poisson distribution (respectively geometric or negative
binomial) with the \(\lambda\) parameter (or \(\lambda,\alpha\)); no
such interpretation exists for hurdle models. Interestingly, estimates
from hurdle zero-truncated and one-inflated zero-truncated models tend
to be quite close to one another, although more rigorous studies are
required to confirm this observation.

\subsection{Fitting method}\label{fitting-method}

As previously noted, the \pkg{singleRcapture} package can be used to
model the (linear) dependence of all parameters on covariates. A
modified IRLS algorithm is employed for this purpose as presented in
Algorithm \ref{algo-estimation}; full details are available in
\cite{VGAM-main}. In order to apply the algorithm, a modified model
matrix \(\boldsymbol{X}_{\text{vlm}}\) is created when the
\code{estimatePopsize} function is called. In the context of the models
implemented in \pkg{singleRcapture}, this matrix can be written as:

\begin{equation}\label{X_vlm-definition}
  \boldsymbol{X}_{\text{vlm}}=
  \begin{pmatrix}
    \boldsymbol{X}_{1} & \boldsymbol{0}&\dotso &\boldsymbol{0}\cr
    \boldsymbol{0}& \boldsymbol{X}_{2} &\dotso &\boldsymbol{0}\cr
    \vdots & \vdots & \ddots & \vdots\cr
    \boldsymbol{0}& \boldsymbol{0}&\dotso &\boldsymbol{X}_{p}
  \end{pmatrix}
\end{equation}

where each \(\boldsymbol{X}_{i}\) corresponds to a model matrix
associated with a user specified formula.

\begin{algorithm}[ht!]
\small
\caption{The modified IRLS algorithm used in the \pkg{singleRcapture} package}
\label{algo-estimation}\DontPrintSemicolon
\nlset{1} Initialize with \code{iter}$\leftarrow 1, \boldsymbol{\eta}\leftarrow$\code{start}
    $, \boldsymbol{W}\leftarrow I, \ell\leftarrow\ell(\boldsymbol{\beta})$.\;
\nlset{2} Store values from the previous step:
    $\ell_{-}\leftarrow\ell, \boldsymbol{W}_{-}\leftarrow\boldsymbol{W}, \boldsymbol{\beta}_{-}\leftarrow\boldsymbol{\beta}$
    (the last assignment is omitted during the first iteration), and assign values in the current iteration
    $\displaystyle\boldsymbol{\eta}\leftarrow\boldsymbol{X}_{\text{vlm}}\boldsymbol{\beta}+\boldsymbol{o}, \boldsymbol{W}_{(k)}\leftarrow\mathbb{E}\left[-\frac{\partial^{2}\ell}{\partial\boldsymbol{\eta}_{(k)}^\top\partial\boldsymbol{\eta}_{(k)}}\right], \boldsymbol{Z}_{(k)}\leftarrow\boldsymbol{\eta}_{(k)}+\frac{\partial\ell}{\partial\boldsymbol{\eta}_{(k)}}\boldsymbol{W}_{(k)}^{-1}-\boldsymbol{o}_{(k)}$,\;
    where $\boldsymbol{o}$ denotes offset.\;
\nlset{3} Assign the current coefficient value:
    $\boldsymbol{\beta}\leftarrow\left(\boldsymbol{X}_{\text{vlm}}\boldsymbol{W}\boldsymbol{X}_{\text{vlm}}\right)^{-1}\boldsymbol{X}_{\text{vlm}}\boldsymbol{W}\boldsymbol{Z}$.\;
\nlset{4} If $\ell(\boldsymbol{\beta})<\ell(\boldsymbol{\beta}_{-})$ try selecting the smallest value $h$ such that for
    $\boldsymbol{\beta}_{h}\leftarrow2^{-h}\left(\boldsymbol{\beta}+\boldsymbol{\beta}_{-}\right)$ the inequality $\ell(\boldsymbol{\beta}_{h})>\ell(\boldsymbol{\beta}_{-})$
    holds if this is successful $\boldsymbol{\beta}\leftarrow\boldsymbol{\beta}_{h}$, else stop the algorithm.\;
\nlset{5} If convergence is achieved or \code{iter} is higher than \code{maxiter}, stop the algorithm,
    else \code{iter}$\leftarrow 1+$\code{iter} and return to step 2.
\end{algorithm}

In the case of multi-parameter families, we get a matrix of linear
predictors \(\boldsymbol{\eta}\) instead of a vector, with the number of
columns matching the number of parameters in the distribution.
``Weights'' (matrix \(\boldsymbol{W}\)) are then modified to be
information matrices
\(\displaystyle\mathbb{E}\left[-\frac{\partial^{2}\ell}{\partial\boldsymbol{\eta}_{(k)}^\top\partial\boldsymbol{\eta}_{(k)}}\right]\),
where \(\ell\) is the log-likelihood function and
\(\boldsymbol{\eta}_{(k)}\) is the \(k\)-th row of
\(\boldsymbol{\eta}\), while in the typical IRLS they are scalars
\(\displaystyle\mathbb{E}\left[-\frac{\partial^{2}\ell}{\partial\eta_{k}^{2}}\right]\),
which is often just
\(\displaystyle-\frac{\partial^{2}\ell}{\partial\eta^{2}}\).

\subsection{Bootstrap variance estimators}\label{sec-boostrap}

We have implemented three types of bootstrap algorithms: parametric
(adapted from theory in \cite{zwane}, \cite{norrpoll} for multiple
source setting with covariates), semi-parametric (see
e.g.~\cite{BoehningFriedl2021}) and nonparametric. The nonparametric
version is the usual bootstrap algorithm; which will typically
underestimate the variance of \(\hat{N}\). In this section, the focus is
on the first two approaches.

The idea of semi-parametric bootstrap is to modify the usual bootstrap
to include the additional uncertainty resulting from the fact that the
sample size is a random variable. This type of bootstrap is performed in
steps listed in Algorithm \ref{algo-semipar-boot}.

\begin{algorithm}[ht!]
\small
\caption{Semi-parametric bootstrap}
\label{algo-semipar-boot}\DontPrintSemicolon
\nlset{1} Draw a sample of size $N_{obs}'\sim\text{Binomial}\left(N', \frac{N_{obs}}{N'}\right)$, where $N'=\lfloor\hat{N}\rfloor+\text{Bernoulli}\left(\lfloor\hat{N}\rfloor-\hat{N}\right)$.\;
\nlset{2} Draw $N_{obs}'$ units from the data uniformly without replacement.\;
\nlset{3} Obtain a new population size estimate $N_b$ using bootstrap data.\;
\nlset{4} Repeat $1-3$ steps $B$ times.
\end{algorithm}

In other words, we first draw a sample size and then a sample
conditional on the sample size. Note that when using the semi-parametric
bootstrap one implicitly assumes that the population size estimate
\(\hat{N}\) is accurate. The last implemented bootstrap type is the
parametric algorithm, which first draws a finite population of size
\(\approx\hat{N}\) from the superpopulation model and then samples from
this population according to the selected model, as described in
Algorithm \ref{algo-par-boot}.

\begin{algorithm}[ht!]
\small
\caption{Parametric bootstrap}
\label{algo-par-boot}\DontPrintSemicolon
\nlset{1} Draw the number of covariates equal to $\lfloor\hat{N}\rfloor+\text{Bernoulli}\left(\lfloor\hat{N}\rfloor-\hat{N}\right)$ proportional to the estimated contribution $(\mathbb{P}\left[Y_{k}>0|\boldsymbol{x}_{k}\right])^{-1}$ with replacement.\;
\nlset{2} Using the fitted model and regression coefficients $\hat{\boldsymbol{\beta}}$ draw for each covariate the $Y$ value from the corresponding probability measure on $\mathbb{N}\cup\{0\}$.\;
\nlset{3} Truncate units with the drawn $Y$ value equal to $0$.\;
\nlset{4} Obtain a population size estimate $N_b$ based on the truncated data.\;
\nlset{5} Repeat $1-4$ steps $B$ times.
\end{algorithm}

Note that in order for this type of algorithm to result in consistent
standard error estimates, it is imperative that the estimated model for
the entire superpopulation probability space is consistent, which may be
much less realistic than in the case of the semi-parametric bootstrap.
The parametric bootstrap algorithm is the default option in
\pkg{singleRcapture}.

\section{The main function}\label{sec-main}

\subsection[The estimatePopsize function]{The \code{estimatePopsize} function}

The \pkg{singleRcapture} package is built around the
\code{estimatePopsize} function. The main design objective was to make
using \code{estimatePopsize} as similar as possible to the standard
\code{stats::glm} function or packages for fitting zero-truncated
regression models, such as \pkg{countreg}
(e.g.~\code{countreg::zerotrunc} function). The \code{estimatePopsize}
function is used to first fit an appropriate (vector) generalized linear
model and to estimate the population size along with its variance. It is
assumed that the response vector (i.e.~the dependent variable)
corresponds to the number of times a given unit was observed in the
source. The most important arguments are given in Table
\ref{tab-arguments-popsize}; the obligatory ones are
\code{formula, data, model}.

\begin{table}[ht!]
\centering
\begin{tabular}{p{3cm}p{11cm}}
\hline
Argument & Description \\
\hline
\code{formula} & The main formula (i.e for the Poisson $\lambda$ parameter); \\
\code{data} & a \code{data.frame} (or \code{data.frame} coercible) object; \\
\code{model} & either a function a string or a family class object specifying which model should be used; possible values are listed in the documentation. The supplied argument should have the form \code{model =  "ztpoisson"}, \code{model = ztpoisson}, or if a link function should be specified, then \code{model = ztpoisson(lambdaLink = "log")} can be used; \\
\code{method} & a numerical method used to fit regression \code{IRLS} or \code{optim}; \\
\code{popVar} & a method for estimating variance of $\hat{N}$ and creating confidence intervals (either bootstrap, analytic or skipping the estimation entirely); \\
\code{controlMethod}, \code{controlModel} or \code{controlPopVar} & control parameters for numerical fitting, specifying additional formulas (inflation, dispersion) and population size estimation, respectively; \\
\code{offset} &  a matrix of offset values with the number of columns matching the number of distribution parameters providing offset values to each of linear predictors;\\
\code{...} & additional optional arguments passed to other methods eg. \code{estimatePopsizeFit}; \\
\hline
\end{tabular}
\caption{A description of \code{estimatePopsize} function arguments}
\label{tab-arguments-popsize}
\end{table}

An important step in using \code{estimatePopsize} is specifying the
\code{model} parameter, which indicates the type of model that will be
used for estimating the \emph{unobserved} part of the population. For
instance, to fit Chao's or Zelterman's model one should select
\code{chao} or \code{zelterman} and, assuming that one-inflation is
present, one can select one of the zero-truncated one-inflated
(\code{ztoi}\(^\ast\)) or one-inflated zero-truncated
(\code{oizt}\(^\ast\)) models, such as \code{oiztpoisson} for Poisson or
\code{ztoinegbin} for NB2.

If it is assumed that heterogeneity is observed for NB2 models, one can
specify the formula in the \code{controlModel} argument with the
\code{controlModel} function and the \code{alphaFormula} argument. This
enables the user to provide a formula for the dispersion parameter in
the NB2 models. If heterogeneity is assumed for \code{ztoi}\(^\ast\) or
\code{oizt}\(^\ast\), one can specify the \code{omegaFormula} argument,
which corresponds to the \(\omega\) parameter in these models. Finally,
if covariates are assumed to be available for the hurdle models
(\code{ztHurdle}\(^\ast\) or \code{Hurdlezt}\(^\ast\)), then
\code{piFormula} can be specified, as it provides a formula for the
probability parameter in these models.

\subsection[Controlling variance estimation with controlPopVar]{Controlling variance estimation with \code{controlPopVar}}

The \code{estimatePopsize} function makes it possible to specify the
variance estimation method via \code{popVar} (e.g.~analytic or variance
bootstrap) and control the estimation process by specifying
\code{controlPopVar}. In the control function \code{controlPopVar} the
user can specify the \code{bootType} argument, which has three possible
values: \code{"parametric", "semiparametric"} and
\code{"nonparametric"}. Additional arguments accepted by the
\code{contorlPopVar} function, which are relevant to bootstrap, include:

\begin{itemize}
  \item \code{alpha}, \code{B} -- the significance level and the number of bootstrap samples to be performed, respectively, with $0.05$ and $500$ being the default options.
  \item \code{cores} -- the number of process cores to be used in bootstrap (1 by default); parallel computing is enabled by \pkg{doParallel} \citep{doParallel}, \pkg{foreach} \citep{foreach} and \pkg{parallel} packages \citep{parallel}.
  \item \code{keepbootStat} -- a logical value indicating whether to keep a vector of statistics produced by the bootstrap.
  \item \code{traceBootstrapSize}, \code{bootstrapVisualTrace} --  logical values indicating whether sample and population size should be tracked (\code{FALSE} by default); these work only when \code{cores} = 1.
    \item \code{fittingMethod}, \code{bootstrapFitcontrol} -- the fitting method (by default the same as the one used in the original call) and control parameters (\code{controlMethod}) for model fitting in the bootstrap.
\end{itemize}

In addition, the user can specify the type of confidence interval by
means of \code{confType} and the type of covariance matrix by using
\code{covType} for the analytical variance estimator (observed or the
Fisher information matrix).

In the next sections we present a case study involving the use of a
simple zero-truncated Poisson regression and a more advanced model:
one-inflated zero-truncated geometric regression with the \code{cloglog}
link function. First, we present the example dataset, then we describe
how to estimate the population size and assess the quality and
diagnostics measures. Finally, we show how to estimate the population
size in user-specified sub-populations.

\section{Data analysis example}\label{sec-study}

The package can be installed in the standard manner using:

\begin{CodeChunk}
\begin{CodeInput}
R> install.packages("singleRcapture")
\end{CodeInput}
\end{CodeChunk}

Then, we need to load the package using the following code:

\begin{CodeChunk}
\begin{CodeInput}
R> library(singleRcapture)
\end{CodeInput}
\end{CodeChunk}

\subsection{Dataset}\label{dataset}

We use a dataset from \cite{ztpoisson}, which contains information about
immigrants in four Dutch cities (Amsterdam, Rotterdam, The Hague and
Utrecht), who were staying in the country without a legal permit in 1995
and appeared in police records for that year. This dataset is included
in the package called \code{netherlandsimmigrant}:

\begin{CodeChunk}
\begin{CodeInput}
R> data(netherlandsimmigrant)
R> head(netherlandsimmigrant)
\end{CodeInput}
\begin{CodeOutput}
  capture gender    age       reason       nation
1       1   male <40yrs Other reason North Africa
2       1   male <40yrs Other reason North Africa
3       1   male <40yrs Other reason North Africa
4       1   male <40yrs Other reason         Asia
5       1   male <40yrs Other reason         Asia
6       2   male <40yrs Other reason North Africa
\end{CodeOutput}
\end{CodeChunk}

The number of times each individual appeared in the records is included
in the \code{capture} variable. The available covariates include
\code{gender, age, reason, nation}; the last two represent the reason
for being captured and the region of the world a given person comes
from:

\begin{CodeChunk}
\begin{CodeInput}
R> summary(netherlandsimmigrant)
\end{CodeInput}
\begin{CodeOutput}
    capture         gender         age                reason
 Min.   :1.000   female: 398   <40yrs:1769   Illegal stay: 259
 1st Qu.:1.000   male  :1482   >40yrs: 111   Other reason:1621
 Median :1.000
 Mean   :1.162
 3rd Qu.:1.000
 Max.   :6.000
                    nation
 American and Australia: 173
 Asia                  : 284
 North Africa          :1023
 Rest of Africa        : 243
 Surinam               :  64
 Turkey                :  93
\end{CodeOutput}
\end{CodeChunk}

One notable characteristic of this dataset is that it contains a
disproportionately large number of individuals who were observed only
once (i.e.~1645).

\begin{CodeChunk}
\begin{CodeInput}
R> table(netherlandsimmigrant$capture)
\end{CodeInput}
\begin{CodeOutput}

   1    2    3    4    5    6
1645  183   37   13    1    1
\end{CodeOutput}
\end{CodeChunk}

The basic syntax of \code{estimatePopsize} is very similar to that of
\code{glm}, the same can be said about the output of the summary method
except for additional results of population size estimates (denoted as
\texttt{Population size estimation results}).

\begin{CodeChunk}
\begin{CodeInput}
R> basicModel <- estimatePopsize(
+   formula = capture ~ gender + age + nation,
+   model   = ztpoisson(),
+   data    = netherlandsimmigrant,
+   controlMethod = controlMethod(silent = TRUE)
+ )
R> summary(basicModel)
\end{CodeInput}
\begin{CodeOutput}

Call:
estimatePopsize.default(formula = capture ~ gender + age + nation,
    data = netherlandsimmigrant, model = ztpoisson(), controlMethod = controlMethod(silent = TRUE))

Pearson Residuals:
     Min.   1st Qu.    Median      Mean   3rd Qu.      Max.
-0.486442 -0.486442 -0.298080  0.002093 -0.209444 13.910844

Coefficients:
-----------------------
For linear predictors associated with: lambda
                     Estimate Std. Error z value  P(>|z|)
(Intercept)           -1.3411     0.2149  -6.241 4.35e-10 ***
gendermale             0.3972     0.1630   2.436 0.014832 *
age>40yrs             -0.9746     0.4082  -2.387 0.016972 *
nationAsia            -1.0926     0.3016  -3.622 0.000292 ***
nationNorth Africa     0.1900     0.1940   0.979 0.327398
nationRest of Africa  -0.9106     0.3008  -3.027 0.002468 **
nationSurinam         -2.3364     1.0136  -2.305 0.021159 *
nationTurkey          -1.6754     0.6028  -2.779 0.005445 **
---
Signif. codes:  0 '***' 0.001 '**' 0.01 '*' 0.05 '.' 0.1 ' ' 1

AIC: 1712.901
BIC: 1757.213
Residual deviance: 1128.553

Log-likelihood: -848.4504 on 1872 Degrees of freedom
Number of iterations: 8
-----------------------
Population size estimation results:
Point estimate 12690.35
Observed proportion: 14.8% (N obs = 1880)
Std. Error 2808.165
95% CI for the population size:
          lowerBound upperBound
normal      7186.449   18194.25
logNormal   8431.277   19718.31
95% CI for the share of observed population:
          lowerBound upperBound
normal     10.332933   26.16035
logNormal   9.534288   22.29793
\end{CodeOutput}
\end{CodeChunk}

The output regarding the population size contains the point estimate,
the observed proportion (based on the input dataset), the standard error
and two confidence intervals: one relating to the point estimate, the
second -- to the observed proportion.

According to this simple model, the population size is about 12,500,
with about 15\% of units observed in the register. The 95\% CI under
normality indicates that the true population size is likely to be
between 7,000-18,000, with about 10-26\% of the target population
observed in the register.

Since there is a reasonable suspicion that the act of observing a unit
in the dataset may lead to undesirable consequences for the person
concerned (in this case, a possible deportation, detention or something
similar). For these reasons, the user may consider one-inflated models,
such as one-inflated zero-truncated geometric model (specified by
\code{oiztgeom} family) and those presented below.

\begin{CodeChunk}
\begin{CodeInput}
R> set.seed(123456)
R> modelInflated <- estimatePopsize(
+     formula = capture ~ nation,
+     model   = oiztgeom(omegaLink = "cloglog"),
+     data    = netherlandsimmigrant,
+     controlModel = controlModel(
+         omegaFormula = ~ gender + age
+     ),
+     popVar = "bootstrap",
+     controlPopVar = controlPopVar(bootType = "semiparametric")
+ )
R> summary(modelInflated)
\end{CodeInput}
\begin{CodeOutput}

Call:
estimatePopsize.default(formula = capture ~ nation, data = netherlandsimmigrant,
    model = oiztgeom(omegaLink = "cloglog"), popVar = "bootstrap",
    controlModel = controlModel(omegaFormula = ~gender + age),
    controlPopVar = controlPopVar(bootType = "semiparametric"))

Pearson Residuals:
    Min.  1st Qu.   Median     Mean  3rd Qu.     Max.
-0.41643 -0.41643 -0.30127  0.00314 -0.18323 13.88376

Coefficients:
-----------------------
For linear predictors associated with: lambda
                     Estimate Std. Error z value  P(>|z|)
(Intercept)           -1.2552     0.2149  -5.840 5.22e-09 ***
nationAsia            -0.8193     0.2544  -3.220  0.00128 **
nationNorth Africa     0.2057     0.1838   1.119  0.26309
nationRest of Africa  -0.6692     0.2548  -2.627  0.00862 **
nationSurinam         -1.5205     0.6271  -2.425  0.01532 *
nationTurkey          -1.1888     0.4343  -2.737  0.00619 **
-----------------------
For linear predictors associated with: omega
            Estimate Std. Error z value  P(>|z|)
(Intercept)  -1.4577     0.3884  -3.753 0.000175 ***
gendermale   -0.8738     0.3602  -2.426 0.015267 *
age>40yrs     1.1745     0.5423   2.166 0.030326 *
---
Signif. codes:  0 '***' 0.001 '**' 0.01 '*' 0.05 '.' 0.1 ' ' 1

AIC: 1677.125
BIC: 1726.976
Residual deviance: 941.5416

Log-likelihood: -829.5625 on 3751 Degrees of freedom
Number of iterations: 10
-----------------------
Population size estimation results:
Point estimate 6699.953
Observed proportion: 28.1% (N obs = 1880)
Boostrap sample skewness: 1.621389
0 skewness is expected for normally distributed variable
---
Bootstrap Std. Error 1719.353
95% CI for the population size:
lowerBound upperBound
  5001.409  11415.969
95% CI for the share of observed population:
lowerBound upperBound
  16.46816   37.58941
\end{CodeOutput}
\end{CodeChunk}

According to this approach, the population size is about 7,000, which is
about 5,000 less than in the case of the naive Poisson approach. A
comparison of AIC and BIC suggests that the one-inflation model fits the
data better with BIC for \code{oiztgeom} 1727 and 1757 for
\code{ztpoisson}.

We can access population size estimates using the following code, which
returns a list with numerical results.

\begin{CodeChunk}
\begin{CodeInput}
R> popSizeEst(basicModel)    # alternative: basicModel$populationSize
\end{CodeInput}
\begin{CodeOutput}
Point estimate: 12690.35
Variance: 7885790
95% confidence intervals:
          lowerBound upperBound
normal      7186.449   18194.25
logNormal   8431.277   19718.31
\end{CodeOutput}
\begin{CodeInput}
R> popSizeEst(modelInflated) # alternative: modelInflated$populationSize
\end{CodeInput}
\begin{CodeOutput}
Point estimate: 6699.953
Variance: 2956175
95% confidence intervals:
lowerBound upperBound
  5001.409  11415.969
\end{CodeOutput}
\end{CodeChunk}

The decision whether to use a zero-truncated Poisson or one-inflated
zero-truncated geometric model should be based on the assessment of the
model and the assumptions regarding the data generation process. One
possible method of selection is based on the likelihood ratio test,
which can be computed quickly and conveniently with the \pkg{lmtest}
(\citet{lmtest}) interface:

\begin{CodeChunk}
\begin{CodeInput}
R> library(lmtest)
\end{CodeInput}
\end{CodeChunk}

\begin{CodeChunk}
\begin{CodeInput}
R> lrtest(basicModel, modelInflated,
+        name = function(x) {
+     if (family(x)$family == "ztpoisson")
+         "Basic model"
+     else "Inflated model"
+ })
\end{CodeInput}
\begin{CodeOutput}
Likelihood ratio test

Model 1: Basic model
Model 2: Inflated model
  #Df  LogLik Df  Chisq Pr(>Chisq)
1   8 -848.45
2   9 -829.56  1 37.776  7.936e-10 ***
---
Signif. codes:  0 '***' 0.001 '**' 0.01 '*' 0.05 '.' 0.1 ' ' 1
\end{CodeOutput}
\end{CodeChunk}

However, the above is not a standard method of model selection in SSCR.
The next sections are dedicated to a detailed description of how to
assess the results using standard statistical tests and diagnostics.

\subsection{Testing marginal
frequencies}\label{testing-marginal-frequencies}

A popular method of testing the model fit in single source
capture-recapture studies consists in comparing the fitted marginal
frequencies
\(\displaystyle\sum_{j=1}^{N_{obs}}\hat{\mathbb{P}}\left[Y_{j}=k|\boldsymbol{x}_{j}, Y_{j} > 0\right]\)
with the observed marginal frequencies
\(\displaystyle\sum_{j=1}^{N}\mathcal{I}_{\{k\}}(Y_{j})=\sum_{j=1}^{N_{obs}}\mathcal{I}_{\{k\}}(Y_{j})\)
for \(k\geq1\). If the fitted model bears sufficient resemblance to the
real data collection process, these quantities should be quite close and
both \(G\) and \(\chi^{2}\) tests can be used to test the statistical
significance of the discrepancy with the following \pkg{singleRcapture}
syntax for the Poisson model (rather poor fit):

\begin{CodeChunk}
\begin{CodeInput}
R> margFreq <- marginalFreq(basicModel)
R> summary(margFreq, df = 1, dropl5 = "group")
\end{CodeInput}
\begin{CodeOutput}
Test for Goodness of fit of a regression model:

                 Test statistics df P(>X^2)
Chi-squared test           50.06  1 1.5e-12
G-test                     34.31  1 4.7e-09

--------------------------------------------------------------
Cells with fitted frequencies of < 5 have been grouped
Names of cells used in calculating test(s) statistic: 1 2 3
\end{CodeOutput}
\end{CodeChunk}

and for the one-inflated model (better fit):

\begin{CodeChunk}
\begin{CodeInput}
R> margFreq_inf <- marginalFreq(modelInflated)
R> summary(margFreq_inf, df = 1, dropl5 = "group")
\end{CodeInput}
\begin{CodeOutput}
Test for Goodness of fit of a regression model:

                 Test statistics df P(>X^2)
Chi-squared test            1.88  1    0.17
G-test                      2.32  1    0.13

--------------------------------------------------------------
Cells with fitted frequencies of < 5 have been grouped
Names of cells used in calculating test(s) statistic: 1 2 3 4
\end{CodeOutput}
\end{CodeChunk}

where the \code{dropl5} argument is used to indicate how to handle cells
with less than \(5\) fitted observations. Note, however, that currently
there is no continuity correction.

\subsection{Diagnostics}\label{diagnostics}

The \code{singleRStaticCountData} class has a \code{plot} method
implementing several types of quick demonstrative plots, such as the
rootogram \citep[cf.][]{rootogram}, for comparing fitted and marginal
frequencies, which can be generated with the following syntax:

\begin{CodeChunk}
\begin{CodeInput}
R> plot(   basicModel, plotType = "rootogram", main = "ZT Poisson model")
R> plot(modelInflated, plotType = "rootogram", main = "OI ZT Geometric model")
\end{CodeInput}
\begin{figure}[ht]

{\centering
\includegraphics[width=7.5cm]{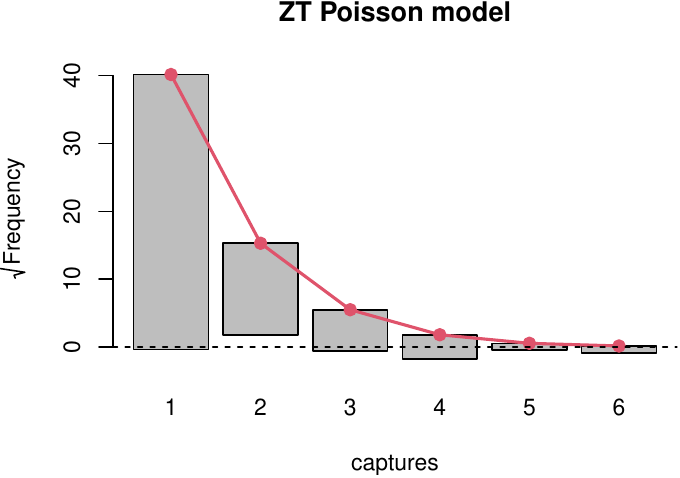}
\includegraphics[width=7.5cm]{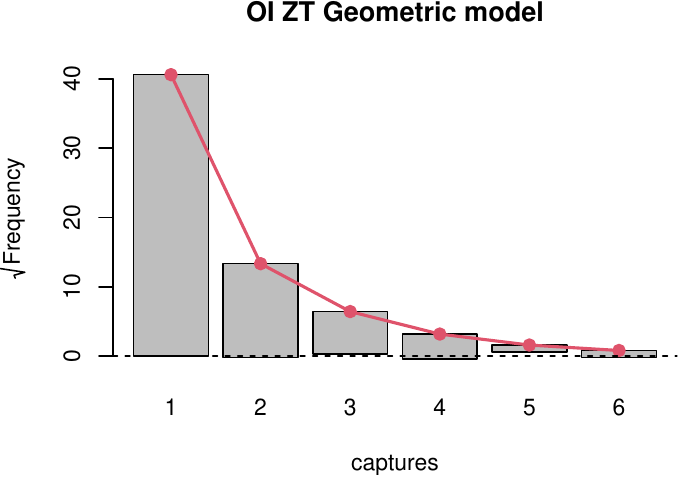}
}

\caption[Rootograms for ztpoisson (left) and oiztgeom (right) models]{Rootograms for ztpoisson (left) and oiztgeom (right) models}\label{fig:rootogram}
\end{figure}
\end{CodeChunk}

The above plots suggest that the \code{oiztgeom} model fits the data
better. Another important issue in population size estimation is to
conduct model diagnostics in order to verify whether influential
observations are present in the data. For this purpose the leave-one-out
(LOO) diagnostic implemented in the \code{dfbeta} from the \pkg{stats}
package has been adapted as shown below (multiplied by a factor of a
hundred for better readability):

\begin{CodeChunk}
\begin{CodeInput}
R> dfb <- dfbeta(basicModel)
R> round(t(apply(dfb, 2, quantile)*100), 4)
\end{CodeInput}
\begin{CodeOutput}
                          0%     25%     50%    75%    100%
(Intercept)          -0.9909 -0.1533  0.0191 0.0521  8.6619
gendermale           -9.0535 -0.0777 -0.0283 0.1017  2.2135
age>40yrs            -2.0010  0.0179  0.0379 0.0691 16.0061
nationAsia           -9.5559 -0.0529  0.0066 0.0120 17.9914
nationNorth Africa   -9.6605 -0.0842 -0.0177 0.0087  3.1260
nationRest of Africa -9.4497 -0.0244  0.0030 0.0083 10.9787
nationSurinam        -9.3138 -0.0065  0.0021 0.0037 99.3383
nationTurkey         -9.6198 -0.0220  0.0079 0.0143 32.0980
\end{CodeOutput}
\end{CodeChunk}

\begin{CodeChunk}
\begin{CodeInput}
R> dfi <- dfbeta(modelInflated)
R> round(t(apply(dfi, 2, quantile)*100), 4)
\end{CodeInput}
\begin{CodeOutput}
                           0%     25%     50%     75%    100%
(Intercept)           -1.4640  0.0050  0.0184  0.0557  9.0600
nationAsia            -6.6331 -0.0346  0.0157  0.0347 12.2406
nationNorth Africa    -7.2770 -0.0768 -0.0170  0.0085  1.9415
nationRest of Africa  -6.6568 -0.0230  0.0081  0.0262  7.1710
nationSurinam         -6.2308 -0.0124  0.0162  0.0421 62.2045
nationTurkey          -6.4795 -0.0273  0.0204  0.0462 21.1338
(Intercept):omega     -6.8668 -0.0193  0.0476  0.0476  9.3389
gendermale:omega      -2.2733 -0.2227  0.1313  0.2482 11.1234
age>40yrs:omega      -30.2130 -0.2247 -0.1312 -0.0663  2.0393
\end{CodeOutput}
\end{CodeChunk}

The result of the \code{dfbeta} can be further used in the
\code{dfpopsize} function, which can be used to quantify LOO on the
population size. Note the warning when the bootstap variance estimation
is applied.

\begin{CodeChunk}
\begin{CodeInput}
R> dfb_pop <- dfpopsize(basicModel, dfbeta = dfb)
R> dfi_pop <- dfpopsize(modelInflated, dfbeta = dfi)
R> summary(dfb_pop)
\end{CodeInput}
\begin{CodeOutput}
     Min.   1st Qu.    Median      Mean   3rd Qu.      Max.
-4236.407     2.660     2.660     5.445    17.281   117.445
\end{CodeOutput}
\begin{CodeInput}
R> summary(dfi_pop)
\end{CodeInput}
\begin{CodeOutput}
     Min.   1st Qu.    Median      Mean   3rd Qu.      Max.
-456.6443   -3.1121   -0.7243    3.4333    5.1535  103.5949
\end{CodeOutput}
\end{CodeChunk}

Figure 2 shows a comparison of the effect of deleting an observation on
the population size estimate and inverse probability weights, which
refer to the contribution of a given observation to the population size
estimate:

\begin{CodeChunk}
\begin{CodeInput}
R> plot(basicModel, plotType = "dfpopContr",
+      dfpop = dfb_pop, xlim = c(-4500, 150))
R> plot(modelInflated, plotType = "dfpopContr",
+      dfpop = dfi_pop, xlim = c(-4500, 150))
\end{CodeInput}
\begin{figure}[ht]

{\centering \includegraphics[width=7.5cm]{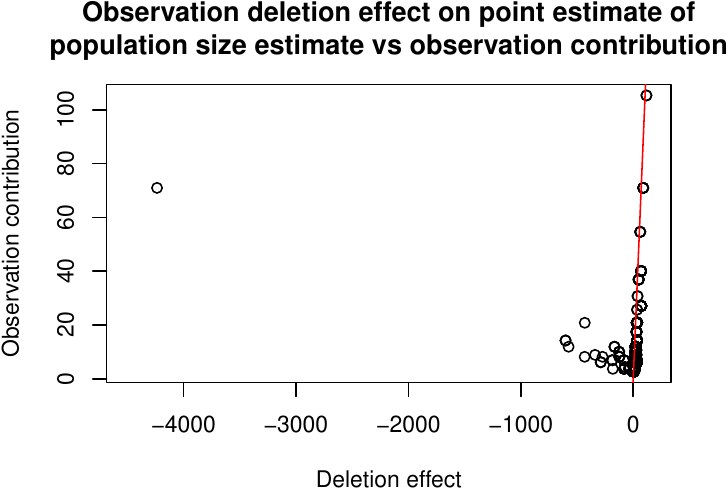} \includegraphics[width=7.5cm]{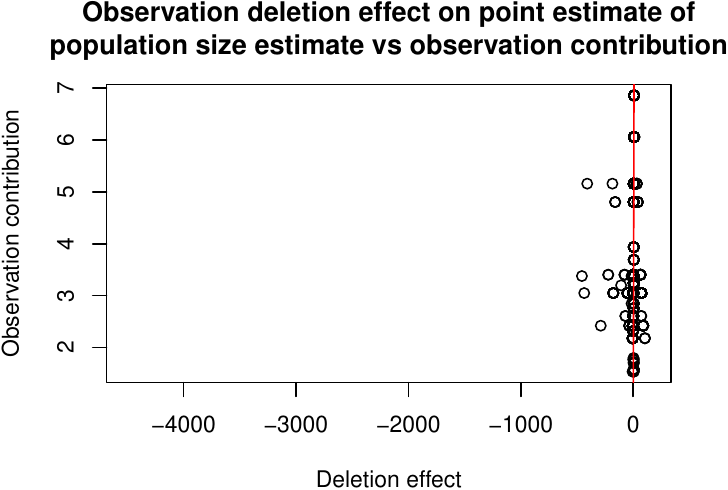}

}

\caption[Results for ztpoisson (left) and oiztgeom (right) model]{Results for ztpoisson (left) and oiztgeom (right) model}\label{fig:dfpopsize_plot}
\end{figure}
\end{CodeChunk}

These plots show how the population size changes if a given observation
is removed. For instance, if we remove observation 542, then the
population size will increase by about 4236 for the \code{ztpoisson}
model. In the case of \code{oiztgeom}, the largest change is equal to
457 for observation 900.

The full list of plot types along with the list of optional arguments
that can be passed from the call to the \code{plot} method down to base
\proglang{R} and \pkg{graphics} functions can be found in the help file
of the \code{plot} method.

\begin{CodeChunk}
\begin{CodeInput}
R> ?plot.singleRStaticCountData
\end{CodeInput}
\end{CodeChunk}

\subsection[The stratifyPopsize function]{The \code{stratifyPopsize} function}

Researchers may be interested not only in the total population size but
also in the size of specific sub-populations (e.g.~males, females,
particular age groups). For this reason we have created the
\code{stratifyPopsize} function, which estimates the size by strata
defined by the coefficients in the model (the default option). The
following output presents results based on the \code{ztpoisson} and
\code{oiztgeom} models.

\begin{CodeChunk}
\begin{CodeInput}
R> popSizestrata <- stratifyPopsize(basicModel)
R> cols <- c("name", "Observed", "Estimated", "logNormalLowerBound",
+           "logNormalUpperBound")
R> popSizestrata_report <- popSizestrata[, cols]
R> cols_custom <- c("Name", "Obs", "Estimated", "LowerBound", "UpperBound")
R> names(popSizestrata_report) <- cols_custom
R> popSizestrata_report
\end{CodeInput}
\begin{CodeOutput}
                             Name  Obs  Estimated LowerBound UpperBound
1                  gender==female  398  3811.0911  2189.0443   6902.133
2                    gender==male 1482  8879.2594  6090.7762  13354.880
3                     age==<40yrs 1769 10506.8971  7359.4155  15426.455
4                     age==>40yrs  111  2183.4535   872.0130   5754.876
5  nation==American and Australia  173   708.3688   504.6086   1037.331
6                    nation==Asia  284  2742.3147  1755.2548   4391.590
7            nation==North Africa 1023  3055.2033  2697.4900   3489.333
8          nation==Rest of Africa  243  2058.1533  1318.7466   3305.786
9                 nation==Surinam   64  2386.4513   505.2457  12287.983
10                 nation==Turkey   93  1739.8592   638.0497   5068.959
\end{CodeOutput}
\end{CodeChunk}

\begin{CodeChunk}
\begin{CodeInput}
R> popSizestrata_inflated <- stratifyPopsize(modelInflated)
R> popSizestrata_inflated_report <- popSizestrata_inflated[, cols]
R> names(popSizestrata_inflated_report) <- cols_custom
R> popSizestrata_inflated_report
\end{CodeInput}
\begin{CodeOutput}
                             Name  Obs Estimated LowerBound UpperBound
1  nation==American and Australia  173  516.2432   370.8463   768.4919
2                    nation==Asia  284 1323.5377   831.1601  2258.9954
3            nation==North Africa 1023 2975.8801  2254.7071  4119.3050
4          nation==Rest of Africa  243 1033.9753   667.6106  1716.4484
5                 nation==Surinam   64  354.2236   193.8891   712.4739
6                  nation==Turkey   93  496.0934   283.1444   947.5309
7                  gender==female  398 1109.7768   778.7197  1728.7066
8                    gender==male 1482 5590.1764  3838.4550  8644.0776
9                     age==<40yrs 1769 6437.8154  4462.3472  9862.2147
10                    age==>40yrs  111  262.1379   170.9490   492.0347
\end{CodeOutput}
\end{CodeChunk}

The \code{stratifyPopsize} function prepared to handle objects of the
\code{singleRStaticCountData} class, accepts three optional parameters
\code{strata, alpha, cov}, which are used for specifying
sub-populations, significance levels and the covariance matrix to be
used for computing standard errors. An example of the full call is
presented below.

\begin{CodeChunk}
\begin{CodeInput}
R> library(sandwich)
R> popSizestrataCustom <- stratifyPopsize(
+   object  = basicModel,
+   strata = ~ gender + age,
+   alpha   = rep(c(0.1, 0.05), each=2),
+   cov     = vcovHC(basicModel, type = "HC4")
+ )
R>
R> popSizestrataCustom_report <- popSizestrataCustom[, c(cols, "confLevel")]
R> names(popSizestrataCustom_report) <- c(cols_custom, "alpha")
R> popSizestrataCustom_report
\end{CodeInput}
\begin{CodeOutput}
            Name  Obs Estimated LowerBound UpperBound alpha
1 gender==female  398  3811.091  2275.6416   6602.161  0.10
2   gender==male 1482  8879.259  6261.5125  12930.751  0.10
3    age==<40yrs 1769 10506.897  7297.2081  15580.138  0.05
4    age==>40yrs  111  2183.453   787.0676   6464.009  0.05
\end{CodeOutput}
\end{CodeChunk}

We have provided integration with the \pkg{sandwich} \citep{sandwich}
package to correct the variance-covariance matrix in the \(\delta\)
method. In the code we have used the \code{vcovHC} method for
\code{singleRStaticCountData} class from the \pkg{sandwich} package,
different significance levels for confidence intervals in each stratum
and a formula to specify that we want estimates for both males and
females to be grouped by \code{nation} and \code{age}. The \code{strata}
parameter can be specified either as:

\begin{itemize}
\item a formula with the empty left hand side, as shown in the example above (e.g. \code{~ gender * age}),
\item a logical vector with the number of entries equal to the number of rows in the dataset, in which case only one stratum will be created (e.g. \code{netherlandsimmigrant$gender == "male"}),
\item a vector of names of explanatory variables, which will result in every level of the explanatory variable having its own sub-population for each variable specified (e.g. \code{c("gender", "age")}),
\item not supplied at all, in which case strata will correspond to levels of each factor in the data without any interactions (string vectors will be converted to factors for the convenience of the user),
\item a (named) list where each element is a logical vector; names of the list will be used to specify variable names in the returned object, for example:
\end{itemize}

\begin{CodeChunk}
\begin{CodeInput}
R> list(
+   "Stratum 1" = netherlandsimmigrant$gender == "male"   &
+     netherlandsimmigrant$nation == "Suriname",
+   "Stratum 2" = netherlandsimmigrant$gender == "female" &
+     netherlandsimmigrant$nation == "North Africa"
+ )
\end{CodeInput}
\end{CodeChunk}

One can also specify \code{plotType = "strata"} in the \code{plot}
function, which results in a plot with point and CI estimates of the
population size.

\begin{CodeChunk}
\begin{CodeInput}
R> par(mar = c(2.5, 8.5, 4.1, 2.5), cex.main = .7, cex.lab = .6)
R> plot(basicModel, plotType = "strata")
R> plot(modelInflated, plotType = "strata")
\end{CodeInput}
\begin{figure}[ht]

{\centering \includegraphics[width=7.5cm]{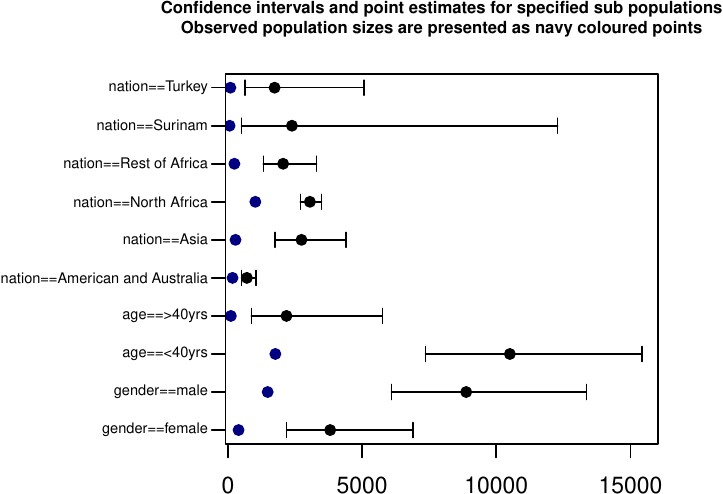} \includegraphics[width=7.5cm]{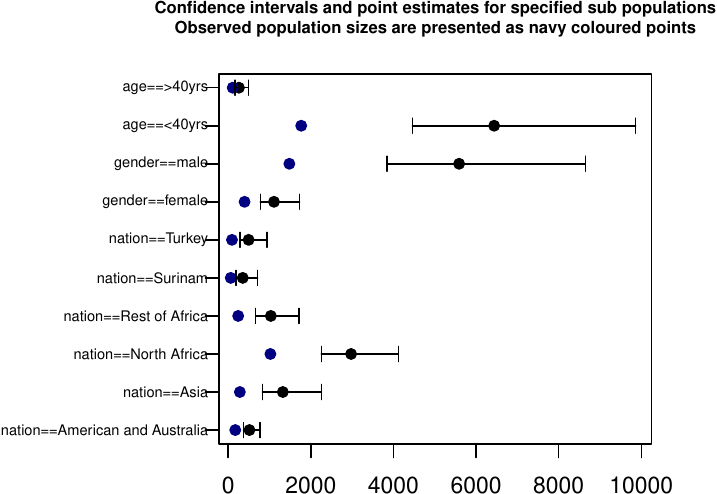}

}

\caption[Population size by covariates for ztpoisson (left) and oiztgeom (right) model]{Population size by covariates for ztpoisson (left) and oiztgeom (right) model}\label{fig:strata_plot}
\end{figure}
\end{CodeChunk}

Only the \code{logNormal} type of confidence interval is used for
plotting since the studentized confidence intervals often result in
negative lower bounds.

\section[Classes and S3Methods]{Classes and \code{S3Methods}}\label{sec-methods}

We have created a number of classes. The main ones are:
\code{singleRStaticCountData}, \code{singleRfamily}, and supplementary
are: \code{popSizeEstResults}, \code{summarysingleRmargin} and
\newline \code{summarysingleRStaticCountData}, which make it possible to
extract relevant information regarding the population size.

For instance, the \code{popSizeEst} function can be used to extract
information about the estimated size of the population as given below:

\begin{CodeChunk}
\begin{CodeInput}
R> (popEst <- popSizeEst(basicModel))
\end{CodeInput}
\begin{CodeOutput}
Point estimate: 12690.35
Variance: 7885790
95% confidence intervals:
          lowerBound upperBound
normal      7186.449   18194.25
logNormal   8431.277   19718.31
\end{CodeOutput}
\end{CodeChunk}

and the resulting object \code{popEst} of the \code{popSizeEstResults}
class contains the following fields:

\begin{itemize}
  \item \code{pointEstimate}, \code{variance} -- numerics containing the point estimate and variance of this estimate.
  \item \code{confidenceInterval} -- a \code{data.frame} with confidence intervals.
  \item \code{boot} -- If the bootstrap was performed a numeric vector containing the $\hat{N}$ values from the bootstrap, a character vector with value \code{"No bootstrap performed"} otherwise.
  \item \code{control} -- a \code{controlPopVar} object with controls used to obtain the object.
\end{itemize}

The only explicitly defined method for \code{popSizeEstResults},
\code{summarysingleRmargin} and \code{summarysingleRStaticCountData}
classes is the \code{print} method, but the former one also accepts
\proglang{R} primitives like \code{coef}:

\begin{CodeChunk}
\begin{CodeInput}
R> coef(summary(basicModel))
\end{CodeInput}
\begin{CodeOutput}
                       Estimate Std. Error    z value      P(>|z|)
(Intercept)          -1.3410661  0.2148870 -6.2407965 4.353484e-10
gendermale            0.3971793  0.1630155  2.4364504 1.483220e-02
age>40yrs            -0.9746058  0.4082420 -2.3873235 1.697155e-02
nationAsia           -1.0925990  0.3016259 -3.6223642 2.919228e-04
nationNorth Africa    0.1899980  0.1940007  0.9793677 3.273983e-01
nationRest of Africa -0.9106361  0.3008092 -3.0272880 2.467587e-03
nationSurinam        -2.3363949  1.0135639 -2.3051284 2.115938e-02
nationTurkey         -1.6753917  0.6027744 -2.7794674 5.444812e-03
\end{CodeOutput}
\end{CodeChunk}

analogously to \code{glm} from \pkg{stats}. The \code{singleRfamily}
inherits the \code{family} class from \pkg{stats} and has explicitly
defined \code{print} and \code{simulate} methods. Example usage is
presented below

\begin{table}[ht!]
\centering
\small
\begin{tabular}{p{4cm}p{11cm}}
\hline
Function & Description \\
\hline
\code{fitted} & it works almost exactly like \code{glm} counterparts but returns more information, namely on fitted values for the truncated and non-truncated probability distribution; \\
\code{logLik} & compared to \code{glm} method, it has the possibility of returning not just the value of the fitted log-likelihood but also the entire function (argument \code{type = "function"}) along with two first derivatives (argument \code{deriv = 0:2}); \\
\code{model.matrix} & it has the possibility of returning the $X_{\text{vlm}}$ matrix defined in \ref{X_vlm-definition};\\
\code{simulate} & it calls the \code{simulate} method for the chosen model and fitted $\boldsymbol{\eta}$; \\
\code{predict} &  it has the possibility of returning either fitted distribution parameters for each unit (\code{type = "response"}), or just linear predictors (\code{type = "link"}), or means of the fitted distributions of $Y$ and $Y|Y>0$ (\code{type = "mean"}) or the inverse probability weights (\code{type = "contr"}). It is possible to set the \code{se.fit} argument to \code{TRUE} in order to obtain standard errors for each of those by using the $\delta$ method. Also, it is possible to use a custom covariance matrix for standard error computation (argument \code{cov}); \\
\code{redoPopEstimation} & a function that applies all post-hoc procedures that were performed (such as heteroscedastic consistent covariance matrix estimation via \pkg{countreg}) to estimate the population size and standard errors; \\
\code{residuals} & used for obtaining residuals of several types, we refer interested readers to the manual \code{?singleRcapture:::residuals.singleRStaticCountData}; \\
\code{stratifyPopsize, summary} & compared to the \code{glm} class, summary has the possibility of adding confidence intervals to the coefficient matrix (argument \code{confint = TRUE}) and using a custom covariance matrix (argument \code{cov = someMatrix}); \\
\code{plot} & it has been discussed above;\\
\code{popSizeEst} & an extractor showcased above; \\
\code{cooks.distance} & it works only for single predictor models \\
\code{dfbeta, dfpopsize} & Multi-threading in \code{dfbeta} is available and \code{dfpopsize} calls \code{dfbeta} if no \code{dfbeta} object was provided in the call; \\
\code{bread, estfun, vcovHC} & for (almost) full \pkg{sandwich} compatibility; \\
\code{AIC, BIC, extractAIC, family, confint, df.residual, model.frame, hatvalues, nobs, print, sigma, influence, rstudent, rstandard}  & it works exactly like \code{glm} counterparts.\\
\hline
\end{tabular}
\caption{\code{S3Methods} implemented in the \pkg{singleRcapture}}
\label{tab-methods}
\end{table}

\begin{CodeChunk}
\begin{CodeInput}
R> set.seed(1234567890)
R> N <- 10000
R> gender <- rbinom(N, 1, 0.2)
R> eta <- -1 + 0.5*gender
R> counts <- simulate(ztpoisson(), eta = cbind(eta), seed = 1)
R> summary(data.frame(gender, eta, counts))
\end{CodeInput}
\begin{CodeOutput}
     gender            eta              counts
 Min.   :0.0000   Min.   :-1.0000   Min.   :0.0000
 1st Qu.:0.0000   1st Qu.:-1.0000   1st Qu.:0.0000
 Median :0.0000   Median :-1.0000   Median :0.0000
 Mean   :0.2036   Mean   :-0.8982   Mean   :0.4196
 3rd Qu.:0.0000   3rd Qu.:-1.0000   3rd Qu.:1.0000
 Max.   :1.0000   Max.   :-0.5000   Max.   :5.0000
\end{CodeOutput}
\end{CodeChunk}

The full list of explicitly defined methods for
\code{singleRStaticCountData} methods is presented in Table
\ref{tab-methods}.

\section{Concluding remarks}\label{concluding-remarks}

In this paper we have introduced the \pkg{singleRcapture} package for
single source capture-recapture models. The package implement
state-of-the-art methods for estimating population size based on a
single data set with multiple counts. The package implements different
methods to account for heterogeneity in capture probabilities, modelled
using covariates, as well as behavioural change, modelled using
one-inflation. We have built the package to facilitate the
implementation of new models using \code{family} objects; their
application is exemplified in the Appendix \ref{sec-family}. An example
of implementing a custom family described in Appendix \ref{sec-custom}
is presented in replication materials.

Furthermore, since many \proglang{R} users are familiar with
\pkg{countreg} or \pkg{VGAM} packages, we have implemented a lightweight
extension called \pkg{singleRcaptureExtra}, available through Github
(\url{https://github.com/ncn-foreigners/singleRcaptureExtra}), which can
be used to integrate \pkg{singleRcapture} with these packages.

In future work we plan to implement Bayesian estimation using
\proglang{Stan} (e.g.~via the \pkg{brms} package;
\cite{carpenter2017stan, brms}) and for one-inflation models we can use
the recent approach proposed by \cite{tuoto2022bayesian} and implement
our own families using the \pkg{brms} package.

\section{Acknowledgements}\label{Acknowledgements}

The authors' work has been financed by the National Science Centre in
Poland, OPUS 20, grant no. 2020/39/B/HS4/00941.

The authors would like to thank Peter van der Heijden, Maarten Cruyff,
Dankmar Böhning and Layna Dennett for useful comments that have helped
to improve the functionality of the package. In addition, we would like
to thank Marcin Szymkowiak and Tymon Świtalski for their valuable
comments that have considerably improved the paper.

\appendix

\section{Detailed information}\label{sec-details}

\subsection[The estimatePopsizeFit function]{The
\code{estimatePopsizeFit} function}\label{sec-estimatePopsizeFit}

In this section we provide a step-by-step description of how to prepare
data in order to use the \code{estimatePopsizeFit} function, which may
be useful to some users, e.g.~those wishing to make modifications to the
\(\hat{N}\) estimate or to the bootstrap. In order to show how to apply
the function we will fit a zero truncated geometric model on the data
from \citet{chao-generalization} with covariate dependency:
\begin{align*}
  \log(\lambda) &=
  \beta_{1, 1} + \beta_{1, 2} \text{log\_distance} + \beta_{1, 3} \text{C\_TYPE} +
  \beta_{1, 4} \text{log\_size}, \\
  \text{logit}(\omega) &=
  \beta_{2, 1} + \beta_{2, 2} \text{log\_distance} + \beta_{2, 3} \text{C\_TYPE}.
\end{align*}

This would be equivalent to the following \code{esimatePopsize} call:

\begin{CodeChunk}
\begin{CodeInput}
R> estimatePopsize(
+   TOTAL_SUB ~ .,
+   data = farmsubmission,
+   model = ztoigeom(),
+   controlModel(
+     omegaFormula = ~ 1 + log_size + C_TYPE
+   )
+ )
\end{CodeInput}
\end{CodeChunk}

\begin{enumerate}
\def\labelenumi{\arabic{enumi}.}
\tightlist
\item
  Create a data matrix \(\boldsymbol{X}_{\text{vlm}}\)
\end{enumerate}

\begin{CodeChunk}
\begin{CodeInput}
R> X <- matrix(data = 0, nrow = 2 * NROW(farmsubmission), ncol = 7)
\end{CodeInput}
\end{CodeChunk}

\begin{enumerate}
\def\labelenumi{\arabic{enumi}.}
\setcounter{enumi}{1}
\tightlist
\item
  Fill the first \(n\) rows with \code{model.matrix} according to the
  specified formula and specify the attribute \code{attr(X, "hwm")} that
  informs the function which elements of the design matrix correspond to
  which linear predictor (covariates for counts and covariates for
  one-inflation)
\end{enumerate}

\begin{CodeChunk}
\begin{CodeInput}
R> X[1:NROW(farmsubmission), 1:4] <- model.matrix(
+   ~ 1 + log_size + log_distance + C_TYPE,
+   farmsubmission
+ )
R> X[-(1:NROW(farmsubmission)), 5:7] <- model.matrix(
+   ~ 1 + log_distance + C_TYPE,
+   farmsubmission
+ )
R> attr(X, "hwm") <- c(4, 3)
\end{CodeInput}
\end{CodeChunk}

\begin{enumerate}
\def\labelenumi{\arabic{enumi}.}
\setcounter{enumi}{2}
\tightlist
\item
  Obtain starting \(\boldsymbol{\beta}\) parameters using the
  \code{glm.fit} function.
\end{enumerate}

\begin{CodeChunk}
\begin{CodeInput}
R> start <- glm.fit(
+   y = farmsubmission$TOTAL_SUB,
+   x = X[1:NROW(farmsubmission), 1:4],
+   family = poisson()
+ )$coefficients
R> start
\end{CodeInput}
\begin{CodeOutput}
[1] -0.82583943  0.33254499 -0.03277732  0.32746933
\end{CodeOutput}
\end{CodeChunk}

\begin{enumerate}
\def\labelenumi{\arabic{enumi}.}
\setcounter{enumi}{3}
\tightlist
\item
  Use the \code{estimatePopsizeFit} function to fit the model assuming a
  zero-truncated one-inflated geometric distribution as specified in the
  \code{family} argument.
\end{enumerate}

\begin{CodeChunk}
\begin{CodeInput}
R> res <- estimatePopsizeFit(
+   y            = farmsubmission$TOTAL_SUB,
+   X            = X,
+   method       = "IRLS",
+   priorWeights = 1,
+   family       = ztoigeom(),
+   control      = controlMethod(silent = TRUE),
+   coefStart    = c(start, 0, 0, 0),
+   etaStart     = matrix(X %*% c(start, 0, 0, 0), ncol = 2),
+   offset       = cbind(rep(0, NROW(farmsubmission)),
+                        rep(0, NROW(farmsubmission)))
+ )
\end{CodeInput}
\end{CodeChunk}

\begin{enumerate}
\def\labelenumi{\arabic{enumi}.}
\setcounter{enumi}{4}
\tightlist
\item
  Compare our results with those obtained by applying the
  \code{stats::optim} function.
\end{enumerate}

\begin{CodeChunk}
\begin{CodeInput}
R> ll <- ztoigeom()$makeMinusLogLike(y = farmsubmission$TOTAL_SUB, X = X)
\end{CodeInput}
\end{CodeChunk}

\begin{CodeChunk}
\begin{CodeInput}
R> res2 <- estimatePopsizeFit(
+   y = farmsubmission$TOTAL_SUB,
+   X = X,
+   method = "optim",
+   priorWeights = 1,
+   family = ztoigeom(),
+   coefStart = c(start, 0, 0, 0),
+   control = controlMethod(silent = TRUE, maxiter = 10000),
+   offset = cbind(rep(0, NROW(farmsubmission)), rep(0, NROW(farmsubmission)))
+ )
\end{CodeInput}
\end{CodeChunk}

\begin{CodeChunk}
\begin{CodeInput}
R> data.frame(IRLS  = round(c(res$beta, -ll(res$beta), res$iter), 4),
+            optim = round(c(res2$beta, -ll(res2$beta), res2$iter[1]), 4))
\end{CodeInput}
\begin{CodeOutput}
         IRLS       optim
1     -2.7845     -2.5971
2      0.6170      0.6163
3     -0.0646     -0.0825
4      0.5346      0.5431
5     -3.1745     -0.1504
6      0.1281     -0.1586
7     -1.0865     -1.0372
8 -17278.7613 -17280.1189
9     15.0000   1696.0000
\end{CodeOutput}
\end{CodeChunk}

The default \code{maxiter} parameter for \code{"optim"} fitting is
\(1000\), but we needed to increase it since the \code{optim} does not
converge in \(1000\) steps and ``gets stuck'' at a plateau, which
results in a lower log-likelihood value compared to the standard
\code{"IRLS"}.

The above situation is rather typical. While we did not conduct any
formal numerical analyses, it seems that when one attempts to model more
than one parameter of the distribution as covariate dependent
\code{optim} algorithms, both \code{"Nelder-Mead"} and \code{"L-BFGS-B"}
seem to be ill-suited for the task despite being provided with the
analytically computed gradient. This is one of the reasons why
\code{"IRLS"} is the default fitting method.

\clearpage

\subsection{Structure of a family function}\label{sec-family}

In this section we provide details regarding the \code{family} object
for the \pkg{singleRcapture} package. This object contains additional
parameters in comparison to the standard \code{family} object from the
\code{stats} package.

\begin{table}[ht!]
\centering
\small
\begin{tabular}{p{4cm}p{11cm}}
\hline
Function & Description \\
\hline
\code{makeMinusLogLike} &
A factory function for creating the following functions:
  \begin{equation*}
    \ell(\boldsymbol{\beta}),
    \frac{\partial\ell}{\partial\boldsymbol{\beta}},
    \frac{\partial^{2}\ell}{\partial\boldsymbol{\beta}^\top\partial\boldsymbol{\beta}}
  \end{equation*}
  from the $\boldsymbol{y}$ vector and the $\boldsymbol{X}_{vlm}$ matrix, which has the \code{deriv} argument with possible
  values in \code{c(0, 1, 2)} that determine which derivative to return; the default value is \code{0}, which represents the minus log-likelihood; \\
\code{links} & A list with link functions; \\
\code{mu.eta, variance} & Functions of linear predictors that return the expected value and variance. The \code{type} argument with 2 possible values (\code{"trunc"} and \code{"nontrunc"}) specifies whether to return $\mathbb{E}[Y|Y>0], \text{var}[Y|Y>0]$ or $\mathbb{E}[Y], \text{var}[Y]$ respectively; the \code{deriv} argument with values in \code{c(0, 1, 2)} is used for indicating the derivative with respect to the linear predictors, which is used for providing standard errors in the \code{predict} method; \\
\code{family} & A string that specifies the model name; \\
\code{valideta, validmu} & For now it only returns \code{TRUE}. In the near future, it will be used to check whether applied linear predictors are valid (i.e. are transformed into some elements of the parameter space subjected to the inverse link function); \\
\code{funcZ, Wfun} & Functions that create pseudo residuals and working weights used in the IRLS algorithm; \\
\code{devResids}  & A function that returns deviance residuals given a vector of prior weights of linear predictors and the response vector; \\
\code{pointEst, popVar} & Functions that return the point estimate for the population size and analytic estimation of its variance given prior weights of linear predictors and, in the later case, also estimates of  $\text{cov}(\hat{\boldsymbol{\beta}})$ and $\boldsymbol{X}_{\text{vlm}}$ matrix. There is an additional boolean parameter \code{contr} in the former function, which, if set to \code{TRUE}, returns the contribution of each unit; \\
\code{etaNames} & Names of linear predictors; \\
\code{densityFunction} & A function that returns the value of PMF at values of \code{x} given linear predictors. The \code{type} argument specifies whether to return $\mathbb{P}[Y|Y>0]$ or $\mathbb{P}[Y]$; \\
\code{simulate} & A function that generates values of a dependent  vector given linear predictors; \\
\code{getStart} & An expression for generating starting points; \\
\hline
\end{tabular}
\end{table}

\clearpage

\section[Implementing custom singleRcapture family function]{Implementing
a custom \pkg{singleRcapture} family function}\label{sec-custom}

Suppose we want to implement a very specific zero truncated family
function in the \pkg{singleRcapture}, which corresponds to the following
``untruncated'' distribution:

\begin{equation}
  \mathbb{P}[Y=y|\lambda, \pi] = \begin{cases}
    1 - \frac{1}{2}\lambda - \frac{1}{2}\pi & \text{when: } y=0\\
    \frac{1}{2}\pi & \text{when: } y=1\\
    \frac{1}{2}\lambda & \text{when: } y=2,
  \end{cases}
\end{equation} with \(\lambda, \pi\in\left(0, 1\right)\) being dependent
on covariates.

In the replication materials we provide a possible way of implementing
the above model, with \code{lambda, pi} meaning
\(\frac{1}{2}\lambda,\frac{1}{2}\pi\). We provide a simple example that
shows that the proposed approach works as expected.

\bibliography{refs.bib}

\end{document}